# PROSPECTION GEOPHYSIQUE DANS LE CADRE D'UNE ETUDE DE LA VARIABILITE SPATIALE DES RENDEMENTS AGRICOLES

**ALBASHA R.[1], THIESSON J.[2], BUVAT S.[2 et 3], LOPEZ J.-M.[4], CHEVIRON B.[1], GUERIN R.[2]**

[1] IRSTEA, UMR G-eau (Gestion de l'Eau, Acteurs, Usages), 361, rue Jean-François Breton BP 5095, F-34196 Montpellier cedex 5, rami.albasha@irstea.fr
[2] Sorbonne Universités, UPMC Univ Paris 06, UMR 7619, METIS, F-75005, Paris, julien.thiesson@upmc.fr
[3] IMG (Ingénierie et Mesures Géophysiques), 28/30 avenue Jacques Anquetil, BP90226, F-95192 Goussainville
[4] CIRAD, UMR G-eau (Gestion de l'Eau, Acteurs, Usages), 361, rue Jean-François Breton BP 5095, F-34196 Montpellier cedex 5

*RESUME*
*Sur le site expérimental de Lavalette (IRSTEA Montpellier) les sources de variabilité du rendement agricole peuvent être l'hétérogénéité spatiale ou temporelle des apports d'eau et d'azote, ainsi que les gradients de propriétés des sols. Ce dernier point est traité en réalisant une prospection géophysique multi-profondeur visant à une cartographie de la résistivité électrique apparente.*
**Mots clés** : *cartographie des sols, résistivité électrique, rendement agricole*

## *GEOPHYSICAL CONTRIBUTION TO STUDY THE SPATIAL VARIABILITY OF AGRICULTURAL YIELDS*

*ABSTRACT*
*Studies conducted on the experimental site of Lavalette (IRSTEA Montpellier) have shown variability in the observed agricultural yield, either attributable to spatial or temporal heterogeneities in water and nitrogen supply or to gradients of soil properties. The latter is addressed by performing a multi-depth geophysical prospection that delivers maps of apparent electrical resistivity.*

## 1. CONTEXTE DE L'ETUDE

Le site expérimental de Lavalette, à la limite nord-est de Montpellier, est utilisé depuis 1989 par l'IRSTEA pour des tests de pratiques agricoles, de stratégies et de techniques d'irrigation (gravitaire, aspersion, goutte à goutte) et de fertigation. Il sert de support à des activités de modélisation et de métrologie des transferts d'eau et d'azote, visant en particulier au développement du modèle PILOTE (MAILHOL et al. 1997, 2011), qui réalise le couplage entre un module hydrologique simplifié et un module de croissance des plantes (maïs sur le site d'étude).

Les terrains se situent dans le lit majeur du Lez, près de sa confluence avec la Lironde, dont le débit est nul en dehors des épisodes pluvieux. La nappe phréatique se trouve habituellement à plusieurs mètres de profondeur, sauf lors des automnes et hivers très pluvieux, où elle peut exceptionnellement affleurer (durant la période de culture de maïs concernée par cette étude, avril-septembre, la nappe se trouve à une profondeur supérieure à 4 m). Le climat est de type méditerranéen et présente un déficit hydro-climatique souvent très prononcé les étés : ETP-Pluie=400 mm de juin à août, en moyenne sur 30 ans. Les sols, très profonds, sont des sols bruns peu évolués, d'origine colluvio-alluviale, à forte réserve utile (environ 180 mm/m), avec très peu d'éléments grossiers, de texture limono-sableuse (au nord de la parcelle, près de la rivière) à limono-argilo-sableuse (au sud de la parcelle, en s'éloignant de la rivière).

Toutefois, les informations disponibles ne sont ni assez détaillées ni suffisamment distribuées spatialement pour obtenir une image fiable de l'hétérogénéité des propriétés du sol, notamment les propriétés pédologiques conditionnant fortement les caractéristiques hydrodynamiques du sol (cette hétérogénéité est suspectée d'être un facteur explicatif pour la variabilité des rendements agricoles observés). La nécessité de trancher cette question impose de mettre en œuvre des méthodes capables de cartographier les gradients (latéraux et verticaux) de propriétés du sol, d'où le choix d'une prospection géophysique multi-profondeur.

## 2. METHODES

Une prospection géophysique par cartographie de la résistivité électrique apparente (appareil RM15, Geoscan Research) avec trois écartements de dispositif pôle-pôle (0.5, 1.0 et 1.5 m) permettant d'atteindre des profondeurs d'investigation de l'ordre de ces écartements, a été réalisée sur le site couvrant une surface d'environ 3600 m$^2$ (Figure 1).

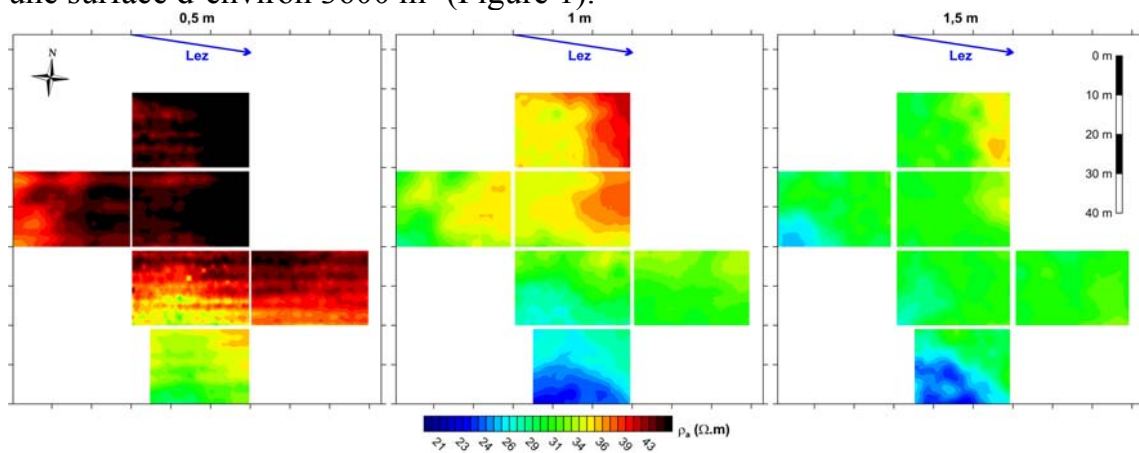

**Fig. 1 – Cartes de résistivité électrique apparente obtenue avec le RM15 (écartement de 0,5, 1 et 1,5 m)**

Nous avons utilisé la méthode de classifications des profils de résistivité développée par BUVAT et al. (2014) afin de produire des cartes donnant des indications sur les changements pédologiques apparaissant latéralement. Cette méthode repose essentiellement sur la définition des taxons (le taxon est le résultat de la classification : tous les profils de résistivités étant classés pour définir des taxons) tels qu'ils apparaissent sur la Figure 2b. L'appartenance à un taxon dépendant du seuil fixé (valeur du paramètre discriminateur α sur la Figure 2a) à partir duquel on considère qu'il y a variation ou non de résistivité apparente avec la profondeur (i.e. suivant les trois écartements).

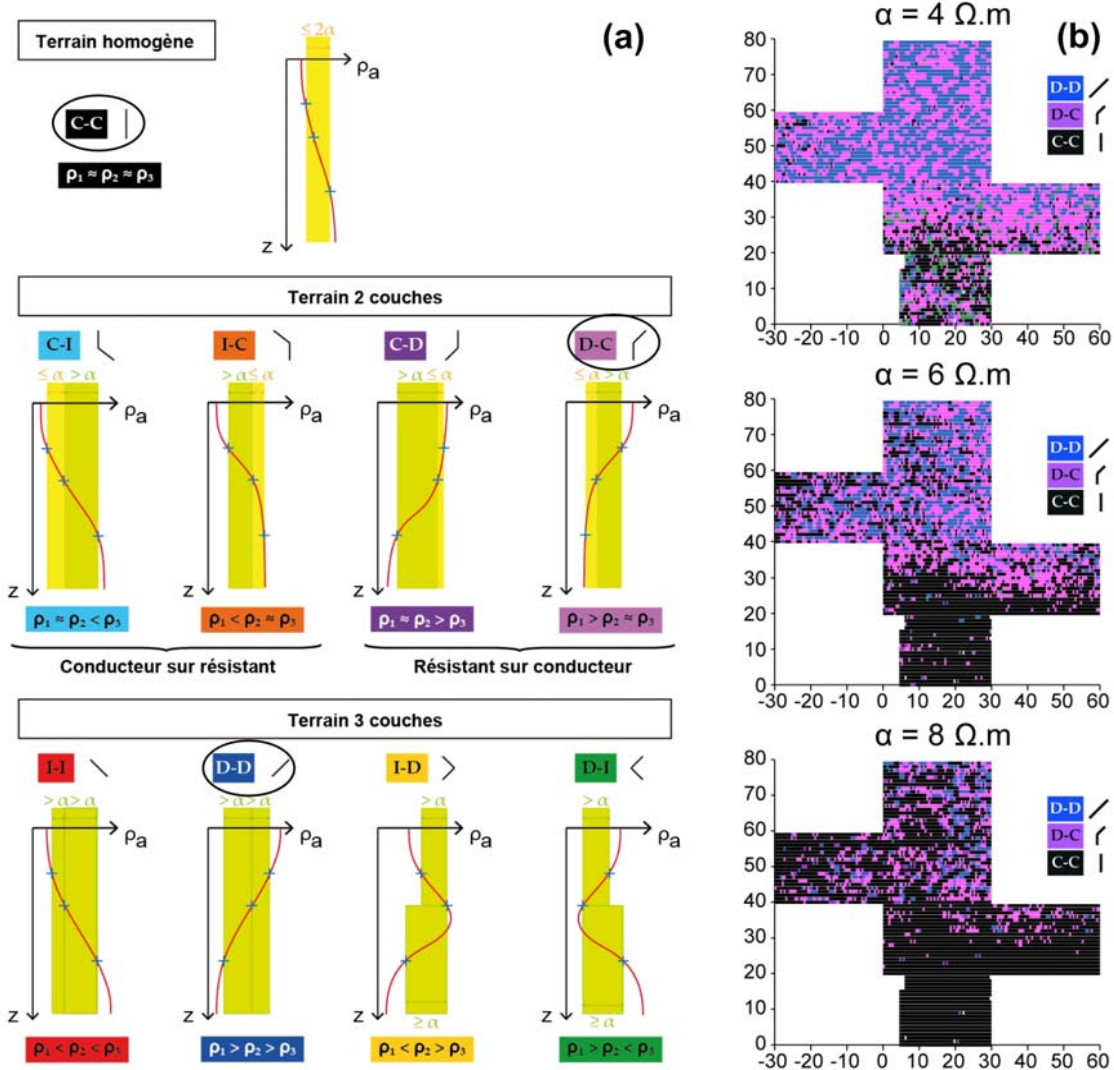

**Fig. 2 – (a) Principe de discrimination des couches, i.e. taxons géophysiques, et (b) Cartographie des profils de résistivité pour différentes valeurs du paramètre discriminateur de couches**

### 3.　RESULTATS

La cartographie de résistivité apparente issue de cette prospection montre une variabilité spatiale, dans l'ensemble, plutôt faible, avec des valeurs comprises entre 20 et 60 Ω.m. Il y a une tendance nord-sud selon laquelle la résistivité semble diminuer plus l'on s'éloigne du Lez. La résistivité apparente diminue globalement avec la profondeur (i.e. les écartements de plus en plus grands).

La Figure 2b présente pour différents seuils les cartes de taxons géophysiques obtenus. On constate que la transition du nord-est au sud-ouest est conservée et qu'elle est constituée principalement d'un épaississement de la couche superficielle plus résistante (passage de D-D à D-C puis à C-C). Ce changement peut être mis en parallèle avec celui qui est observé dans la composition granulométrique du sol superficiel (plus sableux au nord-est qu'au sud-ouest).

## 4. PERSPECTIVES

La validation des interprétations géophysiques présentées passe par une comparaison avec les résultats des carottages prévus prochainement, et avec la carte de la répartition spatiale du rendement du maïs (récolte prévue début août).

Des mesures géophysiques complémentaires sont également programmées.
La prospection des propriétés magnétiques du sol (susceptibilité et viscosité magnétiques) a récemment été mise en œuvre pour tenter de spatialiser les propriétés pédologiques (THIESSON et al. 2011). La viscosité magnétique, liée à l'aimantation rémanente après coupure du champ électromagnétique, a notamment montré une corrélation significative avec la teneur en carbone du sol. Une telle prospection pourrait donc apporter des informations indirectes sur l'un des contrôles du développement végétal, l'azote, à partir du rapport C/N. Ce rapport indique le degré d'évolution de la matière organique et constitue un traceur de la disponibilité de l'azote.
Par ailleurs, la prospection RM15 a été réalisée sur une extension spatiale et à une profondeur limitées. Pour obtenir une visualisation en 3D des sols sur l'ensemble du site, il est envisagé d'acquérir des données sur l'ensemble du site expérimental avec des appareils électromagnétiques (ne nécessitant pas de contact avec le sol) : (i) CMD Mini-Explorer (GF Instruments) qui permet d'atteindre des profondeurs similaires au RM15, (ii) EM31 (Geonics) dont la profondeur d'investigation peut aller jusqu'à 5.5 m.

## RÉFÉRENCES BIBLIOGRAPHIQUES

**BUVAT S., THIESSON J., MICHELIN J., NICOULLAUD B., BOURRENANE H., COQUET Y., TABBAGH A., 2014** – Multi-depth electrical resistivity survey for mapping soil units within two 3-ha plots. *Geoderma*, in press.
**MAILHOL J.C., OLUFAYO, O., RUELLE, P., 1997** - AET and yields assessments based on the LAI simulation. Application to sorghum and sunflower crops. *Agricultural Water Management*, 35, pp.167-182.
**MAILHOL, J.C., RUELLE, P., WALSER, S., SCHUTZE, N., DEJEAN, C., 2011** - Analysis of AET and yield prediction under surface and buried drip irrigation systems using the crop model PILOTE and Hydrus-2D. *Agricultural Water Management*, 98, pp.1033-1044.
**THIESSON J., BUVAT S., SEGER M., GIOT G., BESSON A., TABBAGH A., 2011** – The use of magnetic susceptibility and viscosity measurements as a mapping tool for soil properties: DIGISOIL field results. *EGU general assembly Vienna*.